\documentclass[11pt]{article}
\usepackage{amsmath,latexsym,amssymb,psfrag,subeqn,cancel,subfigure,graphicx}
\usepackage{appendix}
\usepackage{cite}
\usepackage{axodraw}
\numberwithin{equation}{section}
\newcommand{\be}{\begin{equation}}
\newcommand{\ee}{\end{equation}}

\newcommand{\ba}{\begin{array}} 
\newcommand{\ea}{\end{array}}
\newcommand{\bal}{\begin{align}} 
\newcommand{\eal}{\end{align}}
\newcommand{\bqa}{\begin{eqnarray}}
\newcommand{\eqa}{\end{eqnarray}}

\renewcommand{\lim}[2]{\begin{tabular}{c} \vspace{-3mm}lim\\ $_{#1\to #2}$\end{tabular}}


\begin{document} \thispagestyle{empty}

\begin{titlepage} 
\vskip-0.5cm
\begin{flushright}
FERMILAB-PUB-11-614-T
\end{flushright}
\vskip0.2cm

\renewcommand{\thefootnote}{\fnsymbol{footnote}}
\vskip 2cm
\begin{center}  
\vspace{0.3cm} \Large {\sc The Forward-Backward Top Asymmetry in a Singlet extension of the MSSM}
\vspace*{1.5cm}
  
\normalsize  
  
{\bf Alejandro de la Puente~\footnote{adelapue@nd.edu}}

\vskip0.8cm
 {\em Department of Physics, University of Notre Dame, Notre Dame, IN 46556, USA}\\[6pt]
  {\em and}\\[6pt]
 {\em Fermi National Accelerator Laboratory, P.O. Box 500, Batavia, IL 60510, USA}\\[6pt]
\renewcommand{\thefootnote}{\arabic{footnote}}
\setcounter{footnote}{0}

\vskip0.6in \end{center}  
   
\centerline{\large\bf Abstract} 
\vspace{.5cm}

\noindent
The CDF and D$\O$ collaborations have recently reported a large forward-backward asymmetry in the $t\bar{t}$ system which deviates from the next to-leading order QCD standard model prediction. We study the  asymmetry in the $t\bar{t}$ system within the framework of singlet extensions of the Minimal Supersymmetric Standard Model. For this purpose, we introduce non-renormalizable couplings between first and third generation of quarks to scalars.  We analyze two limiting cases of the model, characterized by the size of the supersymmetric mass for the singlet superfield. We study both the small and large limits of this mass parameter. We find that in the region of small singlet supersymmetric mass we can obtain a large asymmetry while being consistent with limits on the $t\bar{t}$ production cross section. These results are also consistent with constraints arising from flavor physics, quark masses and top quark decays.

\vspace{1.5cm} 
\begin{center}  
{\it Dedicated to my dear friend, Leven. If only I can have your spirit.}
\end{center}

\vspace{2mm}

  
\end{titlepage}

\section{Introduction}
The CDF and D$\O$ collaborations have recently reported a new measurement of the inclusive forward-backward top asymmetry. In particular, after unfolding they have found~\cite{Aaltonen:2011kc,Abazov:2011rq}
\begin{eqnarray}
A^{t\bar{t}}_{FB}&=&0.158\pm0.072\pm0.017~(\text{CDF with} ~5.3 ~\text{fb}^{-1}), \\
A^{t\bar{t}}_{FB}&=&0.196\pm0.060^{+0.018}_{-0.026}~(\text{D$\O$ with}~5.4 ~\text{fb}^{-1}),
\end{eqnarray}
which is to be compared to the Standard Model (SM) prediction of $0.058\pm0.009$. Furthermore, the CDF collaboration has measured this asymmetry for different regions of $|\Delta y|$, the difference in the pseudo-rapidities of the top and anti-top quarks,
\begin{eqnarray}
A^{t\bar{t}}_{FB}(|\Delta y|<1)&=&0.026\pm0.118, \\
A^{t\bar{t}}_{FB}(|\Delta y|\ge1)&=&0.611\pm0.256.
\end{eqnarray}
In addition, the CDF collaboration provides a measurement of the asymmetry for two different regions of the $t\bar{t}$ invariant mass distribution:
\begin{eqnarray}
A^{t\bar{t}}_{FB}(M_{t\bar{t}}<450~\text{GeV/c}^{2})&=&-0.116\pm0.153, \\
A^{t\bar{t}}_{FB}(M_{t\bar{t}}\ge450~\text{GeV/c}^{2})&=&~~0.475\pm0.114.
\end{eqnarray}
Equation (1.6) has a significance of 3.1 standard deviations from the SM prediction of $0.088\pm0.013$. The D$\O$ collaboration, however, does not find a significant dependence of $A^{t\bar{t}}$ on either $|\Delta y|$ or $M_{t\bar{t}}$.

The close agreement between the CDF and D$\O$ results on the inclusive asymmetry, serve as motivation for building models beyond the SM that may shed light on possible explanations for the large asymmetry. In this work, we introduce a supersymmetric model to explain the large asymmetry. We use an existing variation of the Next-to-Minimal Supersymmetric Standard Model (NMSSM), known as the singlet extended Minimal Supersymmetric Standard Model (MSSM) or S-MSSM~\cite{CDOP1,CDP2}. The S-MSSM was introduced to provide a more natural solution to the so called little hierarchy problem. In this work, the S-MSSM is extended with dimension-five operators in the superpotential, in order to study their contributions to the forward-backward asymmetry, A$^{t\bar{t}}_{FB}$, and the total cross section, $\sigma^{t\bar{t}}$. Within supersymmetric extensions of the SM, models with R-parity violation can contribute to the asymmetry through a t-channel sparticle exchange in the process $d\bar{d}\to t\bar{t}$~\cite{Cao:2009uz}. In a recent work by Isidori et al.~\cite{Isidori:2011dp}, a simple extension of the SM was introduced which incorporates a light fermion and a scalar with mass above the top mass. The authors found viable regions of parameter space where an asymmetry could be generated from the decays of the new scalars. This extension can be accommodated within the MSSM, where the scalar top can be identified with a right-handed stop and the light neutral fermion with the bino. However, in the MSSM, the bino couples with electroweak strength and a large asymmetry can not be generated. It is argued in~\cite{Isidori:2011dp} that a similar analysis could be carried out within singlet extensions of the MSSM.

Models that incorporate scalars within a variety of representations of the SM gauge group have been extensively studied~\cite{Grinstein:2011yv,Patel:2011eh,Ligeti:2011vt,AguilarSaavedra:2011vw,Gresham:2011pa,Shu:2011au,AguilarSaavedra:2011zy,Nelson:2011us,Babu:2011yw,Cui:2011xy,AguilarSaavedra:2011ug,Vecchi:2011ab,Dorsner:2009mq,Cao:2010zb,Shu:2009xf}. Most recently, Blum et al.~\cite{Blum:2011fa} have argued that only a color-singlet weak doublet with an electroweak-scale mass and a very non-generic flavor structure of Yukawa couplings can enhance the top forward-backward asymmetry while being consistent with the $t\bar{t}$ production cross section and invariant mass distribution. The model presented in this work incorporates a gauge singlet. The asymmetry is mediated by the Higgs mass eigenstates, which can be an admixture of the singlet and the up-type Higgs of the MSSM. We will show that relevant effective couplings in the Lagrangian of $O\left(1\right)$ after electroweak symmetry breaking can enhance the top asymmetry.

 Models that incorporate scalars are not the only route to generating a large asymmetry. Models with exotic gluons~\cite{Krnjaic:2011ub,Gresham:2011pa,AguilarSaavedra:2011ug,Cao:2010zb}, Kaluza-Klein modes within extra dimensional models~\cite{Westhoff:2011ir,Djouadi:2011aj,Bauer:2010iq,Djouadi:2009nb}, and models with new vector bosons~\cite{Haisch:2011up,AguilarSaavedra:2011vw,Gresham:2011pa,Shu:2011au,AguilarSaavedra:2011zy,AguilarSaavedra:2011ug,Cao:2010zb} have also been studied in great detail.

This work is organized as follows: In section two the model is introduced, the Higgs spectrum is reviewed and we show how the observables in the top sector are in great part fixed by electroweak symmetry breaking as well as the Higgs spectrum. In section three, the $p\bar{p}\to t\bar{t}$ differential cross section and forward-backward asymmetry are studied, and in particular the interference between the SM and new physics contributions. Section four outlines the experimental constraints on our model. In section five, results are shown as a function of a few free parameters corresponding to the additional operators contributing to the top observables. Results are summarized in section six and the future outlook is presented. 

\section{Model}
In a recent work~\cite{CDOP1,CDP2} by this author and collaborators a generalization of the NMSSM was studied which was designed to make the solution to the little hierarchy problem more natural within a low energy framework. The model differed from the original NMSSM in that supersymmetric mass terms for both the MSSM Higgs fields and the gauge singlet were introduced. In the following, general aspects of this class of models are reviewed. The superpotential governing these models is given by~\footnote{There is no symmetry that forbids the tadpole term but the non-renormalizable theorem will prevent its generation until SUSY is broken, thus it is assumed to be absent. The $S^{3}$ is no longer required to stabilize the potential and it is taken to zero~\cite{CDOP1}.}
\begin{equation}
W_{\mathrm{S-MSSM}}=W_{\mathrm{Yukawa}}+(\mu+\lambda\hat{S})\hat{H_{u}}\hat{H_{d}}+\frac{\mu_{s}}{2}\hat{S}^{2}.
\end{equation}
The scalar potential, including all the allowed soft SUSY-breaking terms is given by
 \begin{eqnarray}
V&=&(m^{2}_{H_{u}}+|\mu+\lambda S|^{2})|H_{u}|^{2}+(m^{2}_{H_{d}}+|\mu+\lambda S|^{2})|H_{d}|^{2}+(m_{s}^{2}+\mu_{s}^{2})|S|^{2} \nonumber \\ 
&+&[B_{s}S^{2}+(\lambda\mu_{s}S^{\dagger}+B_{\mu}+\lambda A_{\lambda}S)H_{u}H_{d}+h.c.] +\lambda^{2}|H_{u}H_{d}|^{2} \nonumber \\
&+&\frac{1}{8}(g^{2}+g'^{2})(|H_{u}|^{2}-|H_{d}|^{2})^{2}+\frac{1}{2}g^{2}|H_{u}^{\dagger}H_{d}|^{2},
\end{eqnarray}
where $m^{2}_{s}$, $B_{s}$ and $A_{\lambda}$ are the soft breaking contributions associated with the singlet. Minimization of the tree-level scalar potential in absence of CP-violating phases leads to the following three conditions:
\begin{equation}
\frac{1}{2}m_{Z}^{2}=\frac{m_{H_{d}}^{2}-m_{H_{u}}^{2}\tan^{2}2\beta}{\tan^{2}\beta-1}-\mu_{eff}^{2},
\end{equation}
\begin{equation}
\sin2\beta=\frac{2B_{\mu,eff}}{m_{H_{u}}+m_{H_{d}}+2\mu^{2}_{eff}+\lambda^{2}v^{2}},
\end{equation}
\begin{equation}
v_{s}=\frac{\lambda v^{2}}{2}\frac{(\mu_{s}+A_{\lambda})\sin2\beta-2\mu}{\lambda^{2}v^{2}+\mu^{2}_{s}+m^{2}_{s}+2B_{s}},
\end{equation}
where $v_{s}=\left<S\right>$ is the vacuum expectation value (vev) of the singlet field, and $v^{2}=v^{2}_{u}+v^{2}_{d}=(174$ GeV)$^{2}$. Furthermore, the following parameters are defined:
\begin{eqnarray}
\mu_{eff}&=&\mu+\lambda v_{s}, \\
B_{\mu,eff}&=&B_{\mu}+\lambda v_{s}(\mu_{s}+A_{\lambda}).
\end{eqnarray}

As in the NMSSM, this class of models leads to a scalar spectrum consisting of three scalars, two pseudoscalars, and one charged Higgs boson. In~\cite{CDOP1}, the model was analyzed in the limit where $\mu_{s}$ was the largest scale in the Higgs sector. It was found that, in this region of parameter space, the vacuum structure of the model was very similar to that of the MSSM. Furthermore, in the limit of $\mu_{s}\to\infty$, the singlet vev, $v_{s}\to0$, and the singlet could be integrated out supersymmetrically. In the Higgs decoupling limit, only one light scalar identified with the SM-like Higgs boson remained with a mass given by
\begin{equation}
m^{2}_{h^{0}}\approx m^{2}_{Z}\cos^{2}2\beta+\frac{2\lambda v^{2}}{\mu_{s}}\left(2\mu\sin2\beta-A_{\lambda}\sin^{2}2\beta\right).
\end{equation}

In the opposite limit, where $\mu_{s}$ is small, studied in~\cite{CDP2}, the vacuum structure of the model can be substantially different from that of the MSSM. In the Higgs decoupling limit, the spectrum was found to include one scalar identified with the SM-like Higgs boson with mass given by
\begin{equation}
 m^{2}_{h^{0}}\approx m^{2}_{Z}\cos^{2}2\beta+\lambda^{2}v^{2}\sin^{2}2\beta-\frac{(m^{2}_{Z}-\lambda^{2}v^{2})^{2}}{m^{2}_{A}}\sin^{2}2\beta\cos^{2}2\beta,
 \end{equation}
where $m^{2}_{A}\approx 2B_{\mu,eff}/\sin2\beta$. On the other hand, there are two lighter mostly-singlet states with masses given by 
 \begin{eqnarray}
  m^{2}_{A_{s}}&\approx&\mu_{s}^{2}+\lambda^{2}v^{2}-\frac{\lambda^{2}v^{2}A^{2}_{\lambda}}{m^{2}_{A}}, \nonumber \\
 m^{2}_{h_{s}}&\approx&\mu_{s}^{2}+\lambda^{2}v^{2}-\frac{\lambda^{2}v^{2}A^{2}_{\lambda}}{m^{2}_{A}} \cos^{2}2\beta.
 \end{eqnarray}
Within the minimal incarnation of this class of models, there is no significant contribution from the Higgs spectrum to $q\bar{q}$ scattering. Therefore, in this work we consider a simple extension of this scenario by introducing the following dimension-five operators in the superpotential:
 \begin{equation}
 W=W_{\mathrm{S-MSSM}}+\frac{\Lambda_{ij}}{M}\hat{S}\hat{H_{u}}\hat{u^{c}_{i}} \hat{Q}_{j}-\frac{\Sigma_{ij}}{M}\hat{S}\hat{H_{d}}\hat{d^{c}_{i}}\hat{Q}_{j}.
 \end{equation}
These interactions allow for t-channel contributions to $q\bar{q}$ scattering mediated by Higgs particles. In particular, off-diagonal elements coupling first and third families will be relevant in generating the forward-backward asymmetry of the $t\bar{t}$ pair. The scale $M$ dictates where these operators arise, and in this work we assume it is not far from the TeV scale. 
 
In light of the results from both the CDF and D$\O$ collaborations, only couplings between first and third generation of quarks will be considered. We will assume a fermion basis where all the SM up-type Yukawa couplings are diagonal before electroweak symmetry breaking. In such a basis we consider the following structure for the $\Lambda$ matrix:
\begin{equation}
\Lambda = \begin{pmatrix} 
0 & 0 & \Lambda_{13} \\ 
0 & 0& 0 \\ 
\Lambda_{31} & 0 & 0
\end{pmatrix} ~.
\end{equation}
Furthermore, we assume that $\Sigma_{ij}\approx0$, effectively yielding no new physics contributions to the top forward-backward asymmetry. At any rate, compared to the $\Lambda$ effects, the corrections from the $\Sigma$ couplings are suppressed since these enter in the asymmetry and cross section through $d\bar{d}$ scattering.

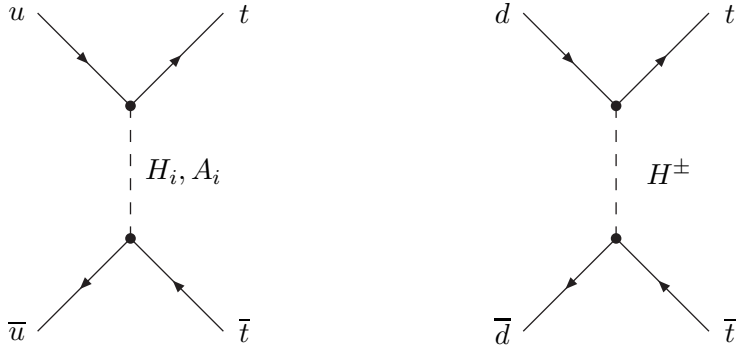
\begin{figure}[ht]
\begin{center}
\begin{picture}(180,120)(-80,-80)
\Text(-43,35)[!]{$u$}
\ArrowLine(-35,35)(0,0)
\ArrowLine(0,0)(35,35)
\Text(43,35)[!]{$t$}
\Vertex(0,0){2}
\Text(20,-25)[!]{$H_{i}, A_{i}$}
\DashLine(0,-50)(0,0){5}
\Vertex(0,-50){2}
\Text(-43,-85)[!]{$\overline{u}$}
\ArrowLine(0,-50)(-35,-85)
\ArrowLine(35,-85)(0,-50)
\Text(43,-85)[!]{$\overline{t}$}
\end{picture}
%
%
\begin{picture}(180,120)(-80,-80)
\Text(-43,35)[!]{$d$}
\ArrowLine(-35,35)(0,0)
\ArrowLine(0,0)(35,35)
\Text(43,35)[!]{$t$}
\Vertex(0,0){2}
\Text(20,-25)[!]{$H^\pm$}
\DashLine(0,-50)(0,0){5}
\Vertex(0,-50){2}
\Text(-43,-85)[!]{$\overline{d}$}
\ArrowLine(0,-50)(-35,-85)
\ArrowLine(35,-85)(0,-50)
\Text(43,-85)[!]{$\overline{t}$}
\end{picture}
\end{center}
\caption{\small New diagrams contributing to $t\bar{t}$ production \label{fig:feynmandiag}}
\end{figure}

The operators in the Lagrangian, derived from (2.12), that couple first generation up quarks to their third generation counterparts through the exchange of neutral scalar or pseudoscalar Higgs bosons are given by
 \begin{equation}
{\cal L}_{u,t}\supset \sum_{i}\left( F^{i}_{R,H}H_{i}-iF^{i}_{R,A}A_{i}\right)\bar{u}_{L}t_{R}+\left( F^{i}_{L,H}H_{i}+iF^{i}_{L,A}A_{i}\right)\bar{u}_{R}t_{L}+h.c.~,
\end{equation} 
where
\begin{eqnarray}
F^{i}_{R,(H,A)}=\frac{\Lambda_{31}}{\sqrt{2}M}(v\sin\beta~O^{(H,A)}_{i,S}+v_{s}~O^{(H,A)}_{i,H_{u}}), \nonumber \\
F^{i}_{L,(H,A)}=\frac{\Lambda_{13}}{\sqrt{2}M}(v\sin\beta~O^{(H,A)}_{i,S}+v_{s}~O^{(H,A)}_{i,H_{u}}).
\end{eqnarray}
The matrices $O^{(H,A)}$ diagonalize the scalar weak eigenstates $(H_{d},H_{u},S)$ into the corresponding mass eigenstates. These are labeled as $(H_{1},H_{2},H_{3})$ for scalars, and $(A_{1},A_{2})$ for pseudoscalars in order of increasing mass. The operators coupling down quarks to top quarks through the exchange of a charged Higgs boson are given by
\begin{equation}
{\cal L}_{d,t}\supset -\frac{v_{s}}{M}\Lambda_{31}\cos\beta~\bar{d}_{L} t_{R}H^{-}+h.c.~.
\end{equation}
The Feynman diagrams corresponding to the new physics in (2.13) and (2.15) are shown in Figure~\ref{fig:feynmandiag}.
%
%

%
%
We will probe the parameter space of the model mainly as a function of the the singlet's vev, $v_{s}$ and the new couplings $\Lambda_{13}$ and $\Lambda_{31}$, always requiring that their values remain below $4\pi$. Notice however, that for such large values of the couplings, extra contributions coming from higher dimensional operators could be of similar size as those given in Equation (2.11). For simplicity, in this work we restrict ourselves to a dimension-five analysis.
%
%
%
%
%
\section{Differential Cross Section and Asymmetry}
Following the analysis carried out by the authors in~\cite{Cao:2010zb} the differential cross section at the parton level can be written as
\begin{equation}
\frac{d\hat{\sigma}}{d\cos\theta}=M^{SM}+M^{INT}+M^{NP},
\end{equation}
where 
$M_{INT}$ denotes the interference between the SM and contributions arising from the operators given in (2.13) and (2.15), while $M_{SM}$ and $M_{NP}$ denote the contributions solely from the SM and new physics, respectively. In what follows, only the interference between new physics and the leading-order standard model diagrams will be considered; we will not incorporate the interference with the dominant NLO QCD 
%
%
%
corrections.
%
%
%
$M_{SM}$ does include next-to-leading order contributions, and so we define the total new physics contributions by
\begin{equation}
M^{NP}_{total}=M^{NP}+M^{SM~LO,~NP}_{INT}.
\end{equation}
Integrating (3.2) in both the forward and backward regions, one can express the asymmetry simply as:
 \begin{equation}
 A^{total}_{FB}=A^{NP}_{FB}\cdot R+A^{SM}_{FB}\cdot(1-R),
 \end{equation}
 where we have made use of the following definitions:
 \begin{eqnarray}
A^{NP}_{FB}&=&\frac{\sigma^{NP}_{F}-\sigma^{NP}_{B}}{\sigma^{NP}_{F}+\sigma^{NP}_{B}}, \nonumber \\
A^{SM}_{FB}&=&\frac{\sigma^{SM}_{F}-\sigma^{SM}_{B}}{\sigma^{SM}_{F}+\sigma^{SM}_{B}}, \\
R&=&\frac{\sigma^{NP}_{total}}{\sigma^{SM}_{total}+\sigma^{NP}_{total}}. \nonumber
\end{eqnarray}

The new physics contributions to the differential cross section in (3.1) can be calculated from equations (2.13) and (2.15). The new physics t-channel contributions to the $t\bar{t}$ production cross section, originating from a $u\bar{u}$ initial state and mediated by scalar and pseudoscalar particles, are given by
\begin{eqnarray}
M^{NP}(u\bar{u}\to t\bar{t})&=&\frac{\pi\beta_{t}\left(\hat{t}-m^{2}_{t}\right)^{2}}{2(16\pi)^{2}\hat{s}}\sum_{ij}\left[\frac{A^{ij}}{(\hat{t}-m^{2}_{H_{i}}+im_{H_{i}}\Gamma(m_{H_{i}}))(\hat{t}-m^{2}_{H_{j}}-im_{H_{j}}\Gamma(m_{H_{j}}))}\right. \nonumber \\
&+&\frac{B^{ij}}{(\hat{t}-m^{2}_{A_{i}}+im_{A_{i}}\Gamma(m_{A_{i}}))(\hat{t}-m^{2}_{A_{j}}-im_{A_{j}}\Gamma(m_{A_{j}}))} \nonumber \\
&+&\left.\left(\frac{C^{ij}}{(\hat{t}-m^{2}_{H_{i}}+im_{H_{i}}\Gamma(m_{H_{i}}))(\hat{t}-m^{2}_{A_{j}}-im_{A_{j}}\Gamma(m_{A_{j}}))}+h.c.\right)\right],
\end{eqnarray}
where $\beta_{t}=\sqrt{1-\frac{4m^{2}_{t}}{\hat{s}}}$ and the expressions for the coefficients $A^{ij},~B^{ij}$ and $C^{ij}$ are given by
\begin{eqnarray}
A^{ij}&=&\left((F^{i}_{R,H}+F^{i}_{L,H})^{2}(F^{j}_{R,H}+F^{j}_{L,H})^{2}+(F^{i}_{R,H}-F^{i}_{L,H})^{2}(F^{j}_{R,H}-F^{j}_{L,H})^{2}\right. \nonumber \\
&+&\left.2(F^{i2}_{R,H}-F^{i2}_{L,H})(F^{j2}_{R,H}-F^{j2}_{L,H})\right), \nonumber \\
B^{ij}&=&\left((F^{i}_{R,A}+F^{i}_{L,A})^{2}(F^{j}_{R,A}+F^{j}_{L,A})^{2}+(F^{i}_{R,A}-F^{i}_{L,A})^{2}(F^{j}_{R,A}-F^{j}_{L,A})^{2}\right. \nonumber \\
&+&\left.2(F^{i2}_{R,A}-F^{i2}_{L,A})(F^{j2}_{R,A}-F^{j2}_{L,A})\right), \nonumber \\
C^{ij}&=&\left((F^{i}_{R,H}+F^{i}_{L,H})^{2}(F^{j}_{R,A}-F^{j}_{L,A})^{2}+(F^{i}_{R,H}-F^{i}_{L,H})^{2}(F^{j}_{R,A}+F^{j}_{L,A})^{2}\right. \nonumber \\
&+&\left.2(F^{i2}_{R,H}-F^{i2}_{L,H})(F^{j2}_{R,A}-F^{j2}_{L,A})\right).
\end{eqnarray}
The contribution arising from a $d\bar{d}$ initial state is mediated by the charged Higgs scalar and it is given by
\begin{equation}
M^{NP}(d\bar{d}\to t\bar{t})=\frac{\pi\beta_{t}\left(\hat{t}-m^{2}_{t}\right)^{2}}{2(8\pi)^{2}\hat{s}}\frac{F^{4}_{H^{\pm}}}{(\hat{t}-m^{2}_{H^{\pm}})^{2}+m^{2}_{H^{\pm}}\Gamma^{2}(m_{H^{\pm}})},
\end{equation}
where $F_{H^{\pm}}=\frac{v_{s}}{M}\Lambda_{31}$. Finally, the interference between the new physics diagrams with those arising from the leading-order QCD contribution are given by
\begin{eqnarray}
M^{INT}(u\bar{u}\to t\bar{t})&=&\frac{\alpha_{s}\beta_{t}}{36\hat{s}^{2}}\sum_{i}\left(\frac{(F^{i2}_{R,(H,A)}+F^{i2}_{L,(H,A)})(\hat{s}m^{2}_{t}+(\hat{t}-m^{2}_{t})^{2})}{\hat{t}-m^{2}_{(H,A)_{i}}+im_{(H,A)_{i}}\Gamma(m_{(H,A)_{i}})}\right. \nonumber \\
&+&\left.\frac{(F^{i2}_{R,(H,A)}+F^{i2}_{L,(H,A)})(\hat{s}m^{2}_{t}+(\hat{t}-m^{2}_{t})^{2})}{\hat{t}-m^{2}_{(H,A)_{i}}-im_{(H,A)_{i}}\Gamma(m_{(H,A)_{i}})}\right),
\end{eqnarray}
for scalar/pseudoscalar mediation and
\begin{eqnarray}
M^{INT}(d\bar{d}\to t\bar{t})&=&\frac{\alpha_{s}\beta_{t}}{36\hat{s}^{2}}\frac{F^{2}_{H^{\pm}}(\hat{s}m^{2}_{t}+(\hat{t}-m^{2}_{t})^{2})}{\hat{t}-m^{2}_{H^{\pm}}+im_{H^{\pm}}\Gamma(m_{H^{\pm}})} \nonumber \\
&+&\frac{F^{2}_{H^{\pm}}(\hat{s}m^{2}_{t}+(\hat{t}-m^{2}_{t})^{2})}{\hat{t}-m^{2}_{H^{\pm}}-im_{H^{\pm}}\Gamma(m_{H^{\pm}})},
\end{eqnarray}
for charged scalar mediation.

\section{Constraints}
 \subsection{$u-t$ Mass Mixing}
 Assuming the fermion basis and the structure of $\Lambda$ introduced in Section 2, the operators in the Lagrangian coupling first generation to third generation up quarks are given by
 \begin{equation}
 {\cal L}_{ut}\supset \frac{\Lambda_{31}}{M}S H^{0}_{u}\bar{t}_{R}u_{L}+\frac{\Lambda_{13}}{M}SH^{0}_{u}\bar{u}_{R}t_{L}+h.c.~.
 \end{equation}
Expanding around fluctuations from the minima of both the singlet and the up-type neutral Higgs, contributions to the masses of the up and the top quarks arise. In particular, these lead to mixing terms parametrized by the following mass matrix:
 \begin{equation}
M^{2}_{U} = \begin{pmatrix} 
\left(\Lambda_{13}~\frac{v_{s}v_{u}}{M}\right)^{2} & \left(\Lambda_{13}~\frac{v_{s}v_{u}}{M}\right)m_{t,0} & \\ 
 \left(\Lambda_{13}~\frac{v_{s}v_{u}}{M}\right)m_{t,0}  & \left(\Lambda_{31}~\frac{v_{s}v_{u}}{M}\right)^{2} +m^{2}_{t,0} \\ 
\end{pmatrix} ~.
\end{equation}
In the above expression the contribution to the up-quark mass from the Yukawa sector has been taken to zero. Furthermore, we use $m_{t,0}$ to denote the contribution from the Yukawa sector to the top quark mass. For $m_{t,0}\gg\Lambda\frac{v_{s}v\sin\beta}{M}$ the following are good approximations to the masses of the quark mass eigenstates  
\begin{eqnarray}
m^{2}_{u}&\approx&\frac{\left(\Lambda_{31}\Lambda_{13}\frac{v^{2}_{s}v^{2}\sin^{2}\beta}{M^{2}}\right)^{2}}{\left( m^{2}_{t,0}+\left(\Lambda_{13}~\frac{v_{s}v_{u}}{M}\right)^{2}+\left(\Lambda_{31}~\frac{v_{s}v_{u}}{M}\right)^{2}\right)}, \nonumber \\
m^{2}_{t}&\approx&\left( m^{2}_{t,0}+\left(\Lambda_{13}~\frac{v_{s}v_{u}}{M}\right)^{2}+\left(\Lambda_{31}~\frac{v_{s}v_{u}}{M}\right)^{2}\right).
\end{eqnarray}
Within this limit  one can can see that $m_{t,0}\approx m_{t}$. The value of $m_{t,0}$ is then found by imposing that $m_{t}\equiv 172.5$ GeV. Experimental constraints on the mass of the up quark~\cite{PDG} give a range of allowed values for $m_{u}$
\begin{equation}
1.3~\text{MeV}\le m_{u}\le3.1~\text{MeV}.
\end{equation}
Imposing $m_{u}\le3.1$ MeV constrains the product of $\Lambda_{13}\cdot\Lambda_{31}$. One can impose that both couplings be small but that will generate no new physics contributions to the forward-backward top asymmetry. One can then impose that either coupling be small enough to satisfy the constrain in (4.4) while making the other provide the new physics to generate a large asymmetry. In what follows, we will see that flavor constraints will constrain $\Lambda_{13}$ over $\Lambda_{31}$.

\subsection{Meson mixing} 
Due to the flavor mixing structure of the matrix $\Lambda$ introduced in (2.13), contributions to meson mixing will arise.
%
%
%
%
The operators in the Lagrangian  contributing to $K^{0}-\bar{K}^{0}$ mixing are given by
\begin{equation}
{\cal L}_{mixing}\supset-\frac{v_{s}}{M}O^{H^{\pm}}_{22}\bar{d}_{Li}(V^{\dagger}\Lambda)_{ij}u_{Rj}H^{-}+h.c.~,
\end{equation}
where $V$ is the CKM matrix. The above contribution to meson mixing has the same structure as that recently studied in~\cite{Blum:2011fa}. In the model considered here the flavor-changing matrix has an additional suppression given by $\frac{v_{s}}{2M}O^{H^{\pm}}_{22}$, and thus it is constrained such that
\begin{equation}
\frac{1}{32\pi^{2}}\left(\frac{\text{TeV}}{m_{H^{\pm}}}\right)^{2}\sum_{i}F\left(x_{i}\right)\left(V^{\dagger}\Lambda'\right)^{2}_{1i}\left(V^{\dagger}\Lambda'\right)^{*2}_{2i}<10^{-6},
\end{equation}
where $x_{i}=\frac{m^{2}_{u_{i}}}{m^{2}_{H^{\pm}}}$ and $\Lambda'=\frac{v_{s}}{2M}O^{H^{\pm}}_{22}\Lambda$. The loop function $F$ is given by
\begin{equation}
F(x)=\frac{1-x^{2}+2x\log(x)}{(1-x)^{3}}.
\end{equation}
Suppressing contributions to $K^{0}-\bar{K}^{0}$ mixing can be achieved with large charged Higgs masses or in the limit where $\Lambda_{13}\ll1$.

\subsection{New Top decay channels}
In Section 2 we introduced the Higgs spectrum of the S-MSSM. In particular, in the Higgs decoupling limit of the model where $\mu_{s}$ corresponds to the largest scale in the Higgs sector, one light scalar exists and can be identified with the SM-like Higgs. In the small $\mu_{s}$ limit, two additional singlet-like scalars with masses below $100$ GeV are present. Due to the new flavor-changing neutral current operators present in our model, the light scalars contribute to the decay width of the top quark. In particular we have for $m_{\phi_{i}}\le m_{t}$:
\begin{equation}
\Gamma \left(t\to\phi_{i} u\right)=\frac{m_{t}}{32\pi}\left(1-\frac{m^{2}_{\phi_{i}}}{m^{2}_{t}}\right)^{2}\left(F^{i2}_{L}+F^{i2}_{R}\right),
\end{equation}
where $\phi_{i}$ denotes any scalar or pseudoscalar that can be produced by a decaying top. A direct measurement of the top decay width has been carried out recently and yields an upper bound on the total decay width of the top quark of $7.6$ GeV at the $95\%$ confidence level, for a top mass of $172.5$ GeV~\cite{Aaltonen:2010ea}. We incorporate this constraint to place bounds on the allowed size for the couplings $F_{L,R}$.
\subsection{Constraints from single and same-sign top production}
It is also worth mentioning the collider constraints from single top production and same-sign top production that may restrict our parameter space. In particular, from~\cite{AguilarSaavedra:2011zy} we see that the coupling which enters into the cross section for same sign-top production is given by
\begin{equation}
g_{tt}\propto\left(\Lambda_{13}\Lambda_{31}\right)^{2}.
\end{equation}
Therefore, one may suppress any additional contributions to the same-sign top production cross section by suppressing one of the couplings as in the $u-t$ mass mixing constraint.

%
%
%
%
%
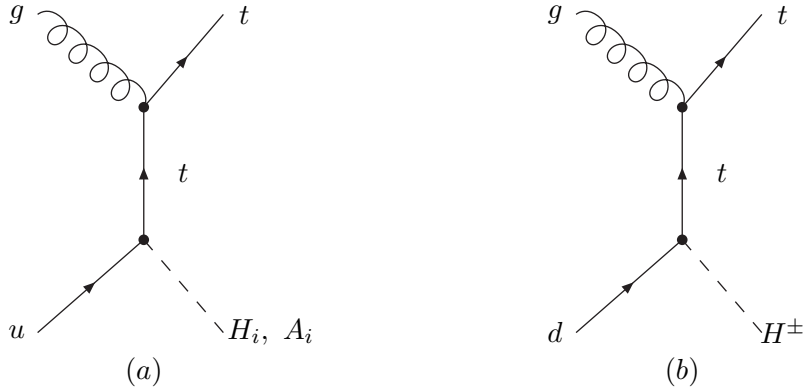
\begin{figure}[t]
\begin{center}
\begin{picture}(150,80)(-50,-100)
\Text(-43,35)[!]{$g$}
\Gluon(-35,35)(5,0){5}{4}
\ArrowLine(5,0)(35,35)
\Text(43,35)[!]{$t$}
\Vertex(5,0){2}
\Text(20,-25)[!]{$t$}
\ArrowLine(5,-50)(5,0)
\Vertex(5,-50){2}
\Text(-43,-85)[!]{$u$}
\ArrowLine(-35,-85)(5,-50)
\DashLine(35,-85)(5,-50){5}
\Text(53,-85)[!]{$H_{i},~A_{i}$}
\Text(5,-100)[!]{$(a)$}
\end{picture}
\begin{picture}(150,80)(-100,-100)
\Text(-43,35)[!]{$g$}
\Gluon(-35,35)(5,0){5}{4}
\ArrowLine(5,0)(35,35)
\Text(43,35)[!]{$t$}
\Vertex(5,0){2}
\Text(20,-25)[!]{$t$}
\ArrowLine(5,-50)(5,0)
\Vertex(5,-50){2}
\Text(-43,-85)[!]{$d$}
\ArrowLine(-35,-85)(5,-50)
\DashLine(35,-85)(5,-50){5}
\Text(43,-85)[!]{$H^\pm$}
\Text(5,-100)[!]{$(b)$}
\end{picture}
\end{center}
\caption{\small New physics diagrams contributing to single top production together with a neutral Higgs in (a) and charged Higgs in (b). \label{fig:singletop}}
\end{figure}
%
%
%
%
The diagrams contributing to single top production are shown in Figure~\ref{fig:singletop}. D$\O$ has a recent model independent measurement on the single top production cross section using center of mass energies of $\sqrt{s}=1.96$ TeV with $5.4$fb$^{-1}$~\cite{Abazov:2011rz}. They find:
\begin{equation}
\sigma\left(p\bar{p}\to tqb+X\right)=2.90\pm0.59~\text{pb},
\end{equation}
for a top mass of $172.5$ GeV. The diagram in Figure~\ref{fig:singletop}b will contribute to the single top production cross section for $H^{\pm}\to\bar{b}u$. This contribution can be suppressed by either suppressing $\Lambda_{13}$ which comes into the $dtH^{\pm}$ vertex or by a suppression of $\Lambda_{31}$ which has the effect of making the decay  $H^{\pm}\to\bar{b}u$ negligible. For a very heavy scalar or pseudoscalar the diagram of Figure~\ref{fig:singletop}a will be naturally suppressed at the Tevatron. For light scalars/pseudoscalars, the only decay channel open is into $b\bar{b}$, in which case the signal will be $t+2b~\text{jets}$. In some cases there are cascade decays between Higgses, which may suppress the branching ratio into $b\bar{b}$ significantly, and the main signal will be $t+4b~\text{jets}$. The coupling at the $tu(H_{i}A_{i})$ vertex is proportional to $F_{L,R}$ in Equation (2.14) and thus one may need to suppress both couplings in order to not enhance the single top production cross section. Given the complexity of the final states, a direct comparison with the D$\O$ measurement is difficult to make.

%
%
%
%
 \section{Results}
 In this section we present results on the forward-backward asymmetry in $t\bar{t}$ production arising from the of the S-MSSM introduced in Section 2, as well as the contributions to the total $t\bar{t}$ cross section. Due to the large number of parameters that are present in our model, and due to the fact that there exists a vast region of parameter space that can provide a solution to the little hierarchy problem, we present our results for various values of $v_{s}$. Furthermore, we illustrate our results for the two limiting cases of $\mu_{s}$ which were explained in Section 2. We use $\tan\beta=2$ with a corresponding value of $\lambda=0.63$~\cite{CDOP1,CDP2} and work in the Higgs decoupling limit  in order to maximize the tree-level contribution to the mass of the SM-like Higgs boson. We fix the scale of new physics, $M$, where the operators in (2.13) arise, to $1$ TeV.
 
 In our calculations we make use the CTEQ6L PDF set~\cite{Pumplin:2002vw} using a factorization and renormalization scale of $m_{t}/2$. For the strong coupling constant we take $\alpha_{s}(161.9~\text{GeV})\sim 0.1$, which is used to calculate the one-loop radiative correction to the Higgs masses in~\cite{CDOP1,CDP2}. We  carry out our calculations using a top mass of $172.5$ GeV and use the CDF analysis of the $t\bar{t}$ production cross section which incorporates a combination of leptonic and hadronic channels using data with an integrated luminosity of $4.6$ fb$^{-1}$~\cite{ttbarCSCDF}. They find:
 \begin{equation}
 \sigma_{t\bar{t}}=7.50\pm0.48~\text{pb},
 \end{equation}
 for $m_{t}=172.5$ GeV. In addition, we apply all of the constraints introduced in Section 4 to search for viable scenarios consistent with experimental observations.
 \begin{figure}[t] \centering
\includegraphics[width=4in]{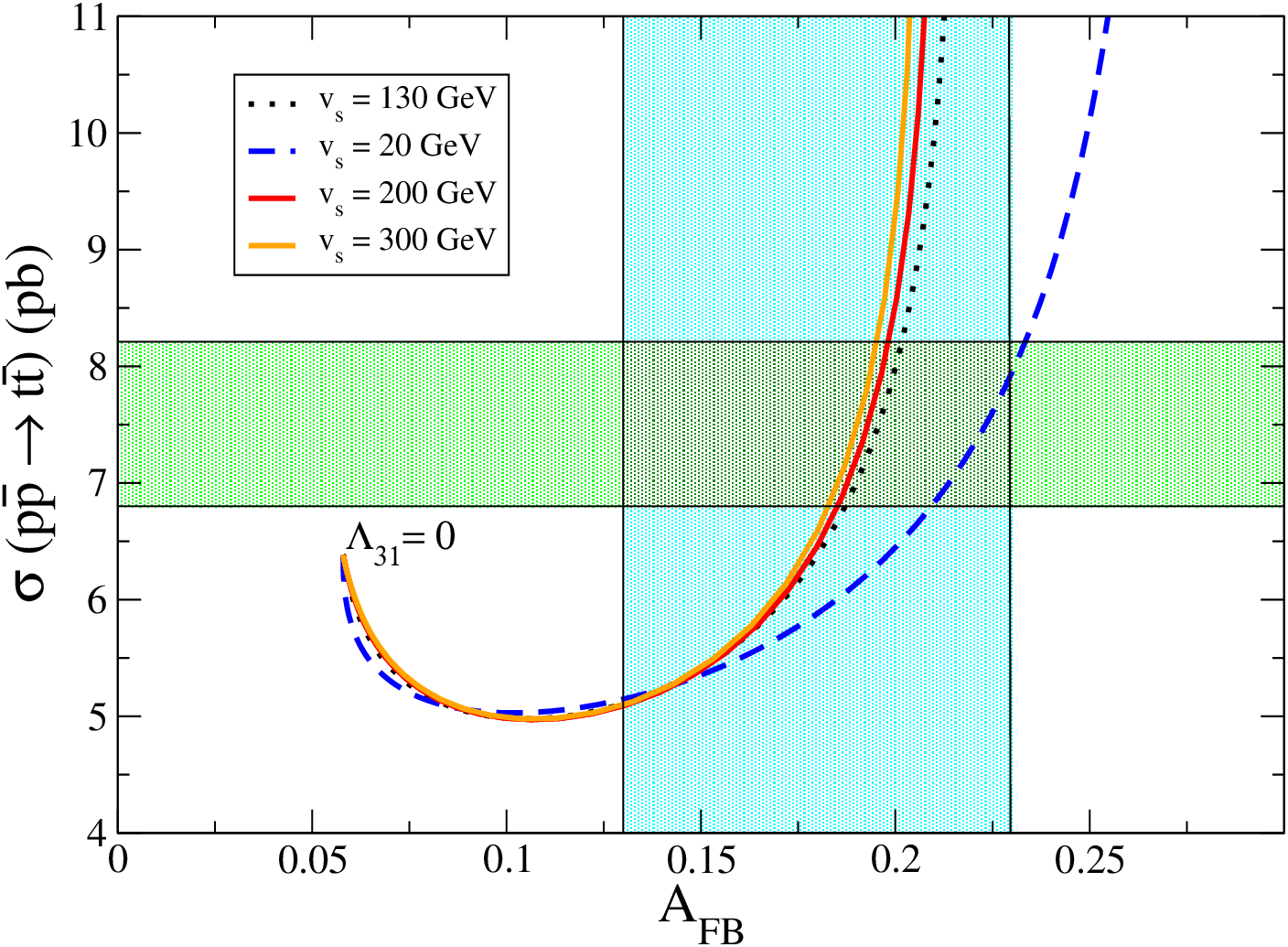} ~~~
\caption{\small The $t\bar{t}$ production cross section as a function of the parton level forward-backward top asymmetry for various values of the singlet vev $v_{s}$, Scenarios A through D. The green band indicate the combined uncertainty from the asymmetry measurements of CDF and D$\O$~\cite{Aaltonen:2011kc,Abazov:2011rq}, and the cyan band the combined theoretical and experimental uncertainty on the value of the $t\bar{t}$ production cross section given in Equation (5.1)~\cite{ttbarCSCDF}. The value of $\Lambda_{31}$ increases along the curves, from $0$ (left) to $9.5$ (right) for $\Lambda_{13}$ close to zero. \label{fig:smallmus1}
}
\end{figure}
 
\begin{figure}[t] 
\begin{center}
\includegraphics[width=2.8in]{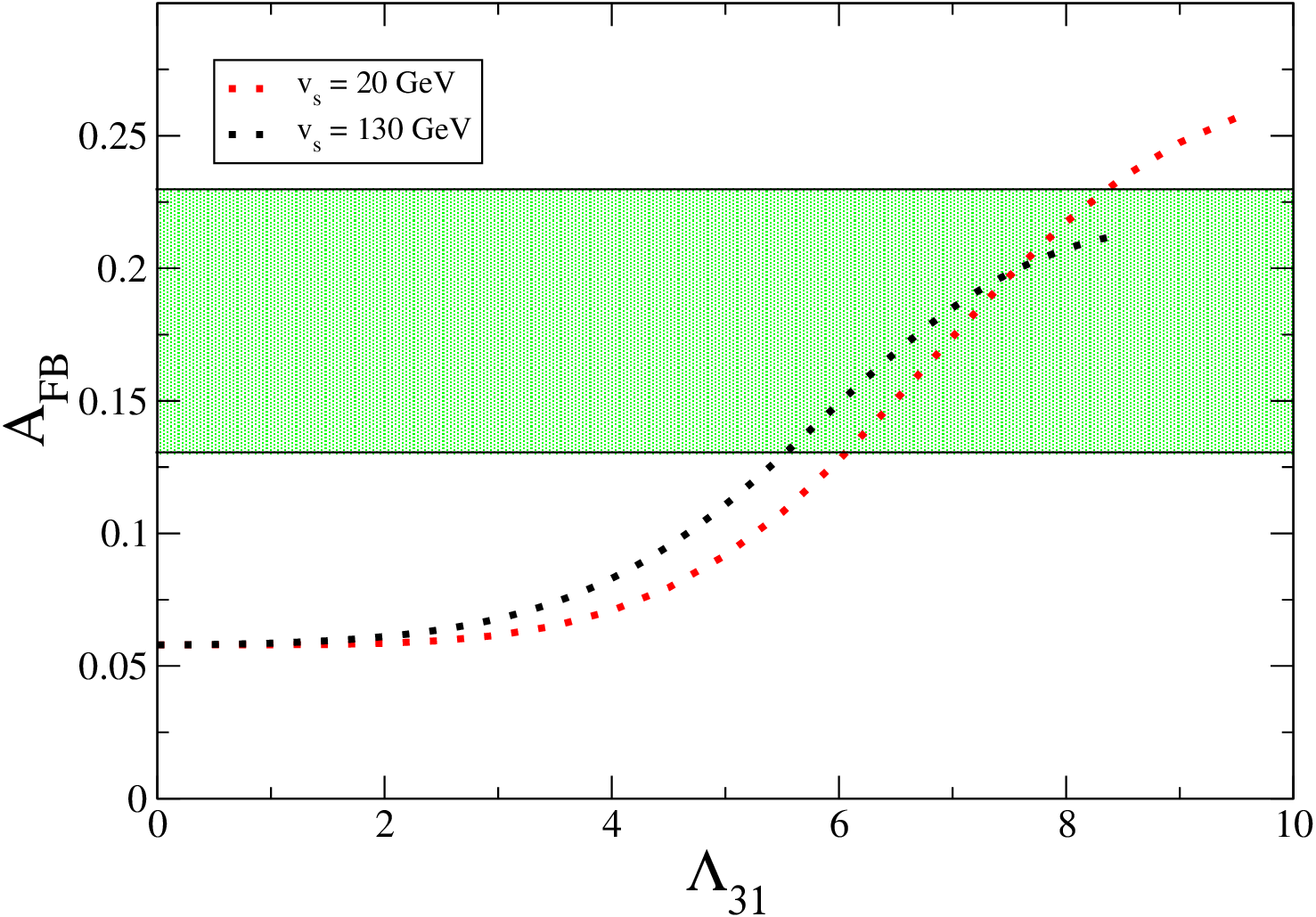}~~
\includegraphics[width=2.8in]{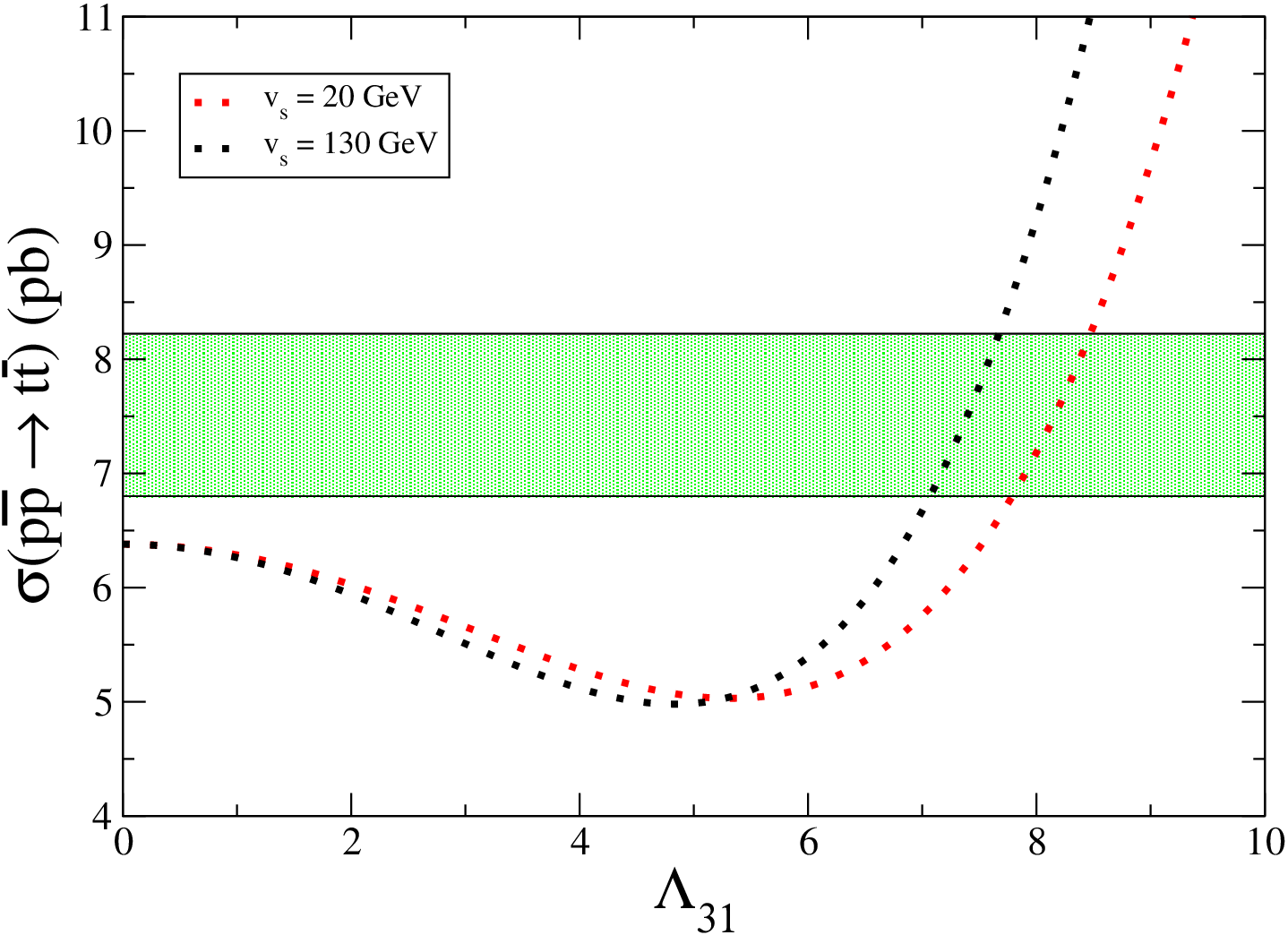}
\caption{\small On the left plot the forward-backward top asymmetry at the parton level as a function of $\Lambda_{31}$ for scenarios A and B. The green bands indicate the combined uncertainty from the asymmetry measurements of CDF and D$\O$~\cite{Aaltonen:2011kc,Abazov:2011rq}. On the right, the $t\bar{t}$ production cross section as a function of $\Lambda_{31}$ for scenarios A and B. The green bands indicate the combined theoretical and experimental uncertainty on the cross section~\cite{ttbarCSCDF}.\label{fig:smallmus2}
}
\end{center}
\end{figure}

 In the small $\mu_{s}$ limit the vacuum structure of the theory is significantly different from that of the MSSM. In particular, the appearance of light mostly singlet scalars can significantly enhance the $t\bar{t}$ production cross section. In Figure~\ref{fig:smallmus1} we illustrate our results of the new physics contributions to the $t\bar{t}$ cross section as a function of the forward-backward asymmetry. The output parameters that arise from electroweak symmetry breaking are shown in Table~\ref{tab:scenarios_input}. In the figure we show the experimental value for the cross section with a green one sigma band and the experimental value for the asymmetry with a cyan one sigma band. As can be seen from the figure, for all of our curves there is a region that falls within one standard deviation from both cross section and asymmetry. The black dotted line corresponds to a value of $v_{S}=120$ GeV, $\mu_{s}=20$ GeV and $A_{\lambda}=190$ GeV as well as vanishing values  of $\mu$ and $B_{\mu}$, labeled scenario A in Table~\ref{tab:scenarios_input}. Scenario A is characterized by a heavy scalar and pseudoscalar with masses around 200 GeV, a SM-like Higgs with mass 124 GeV, one singlet-like scalar with mass 85 GeV and one singlet-like pseudoscalar with a mass of 60 GeV. The mass splitting between the two singlet-like states is evident from Equation (2.10) and it is due to the fact  that the ratio $A^{2}_{\lambda}/m^{2}_{A}$ approaches unity. The blue dotted line corresponds to a value of $v_{S}=20$ GeV, $\mu_{s}=20$ GeV and $A_{\lambda}=470$ GeV as well as values  for $\mu$ and $\sqrt{B_{\mu}}$ of 180 and 500 GeV respectively, labeled scenario B in Table~\ref{tab:scenarios_input}. Scenario B is characterized by a  heavy scalar and pseudoscalar with masses around 800 GeV, a SM-like Higgs with mass 124 GeV, one singlet-like scalar and pseudoscalar with masses close to~$\sim100$ GeV. The near mass degeneracy of the singlet-like states is apparent from Equation (2.10) given that the $A^{2}_{\lambda}/m^{2}_{A}$ ratio has a more negligible contribution to the masses. In Figure~\ref{fig:smallmus2} we plot the asymmetry as a function of $\Lambda_{31}$ on the left, and the total cross section as function of $\Lambda_{31}$ on the right for scenarios A and B. The value of $\Lambda_{13}$ is fixed close to zero in order to remain consistent mainly with the constraint arising from the up quark mass. In this figure, the impact that the lighter spectrum has on the cross section becomes more evident and they become more dominant in scenario A for smaller values of $\Lambda_{31}$. From Figures~\ref{fig:smallmus1} and~\ref{fig:smallmus2} one can also note the inflection point where the pure new physics contributions to the cross section dominate over the interference terms in (3.10) and (3.11). This transition from negative to positive contributions to the cross sections is more rapid for smaller values of $\Lambda_{31}$ and larger values of $v_{s}$, and it is also a consequence of the relatively light spectrum. In scenarios C and D (red and orange in Figure~\ref{fig:smallmus1}, respectively) the value of $v_{s}$ is increased by increasing $A_{\lambda}$ to $310$ and $470$ GeV, respectively. The values of $\mu$ and $B_{\mu}$ are fixed to zero. The light Higgs spectrum for these two scenarios remains identical to that of scenario A, since the ratio of $A^{2}_{\lambda}/m^{2}_{A}$ remains close to unity. A large value of $v_{s}$ thus requires a smaller value of $\Lambda_{31}$ to generate a significant contribution to the cross section. 
%
%
%
%
\begin{table}[hb] 
\addtolength{\arraycolsep}{10pt}
\renewcommand{\arraystretch}{1.3}
\centering
\begin{tabular}{|c|c|c|c|c|c|}
\hline\hline
& Sc. A & Sc. B & Sc. C & Sc. D\\ 
\hline\hline
$v_{s}$ [GeV] & $130$ & $20$ & $200$ & $300$ \\ 
$O^{H}_{2,S}, O^{H}_{2,H_{u}}$ & $-0.079$,~$0.90$ & $0.024$,~$-0.89$ & $-0.12$,~$0.90$ & $-0.18$,~$0.89$\\ 
$O^{H}_{1,S}, O^{H}_{1,H_{u}}$ & $-0.091$,~$0.93$ & $-0.0007$,~$0.99$ & $0.01$,~$0.97$ & $0.10$,~$0.97$ \\ 
$O^{A}_{1,S}, O^{A}_{1,H_{u}}$ & $-0.19$~$0.90$ & $-0.03$~$0.99$ & $-0.13$,~$0.95$ & $-0.095$,~$0.98$\\ 
\hline\hline
\end{tabular}
\caption{\small Scalar mixing angles and vev in the singlet field direction.\label{tab:scenarios_input}
}
\end{table}

\begin{figure}[t] \centering
\includegraphics[width=2.8in]{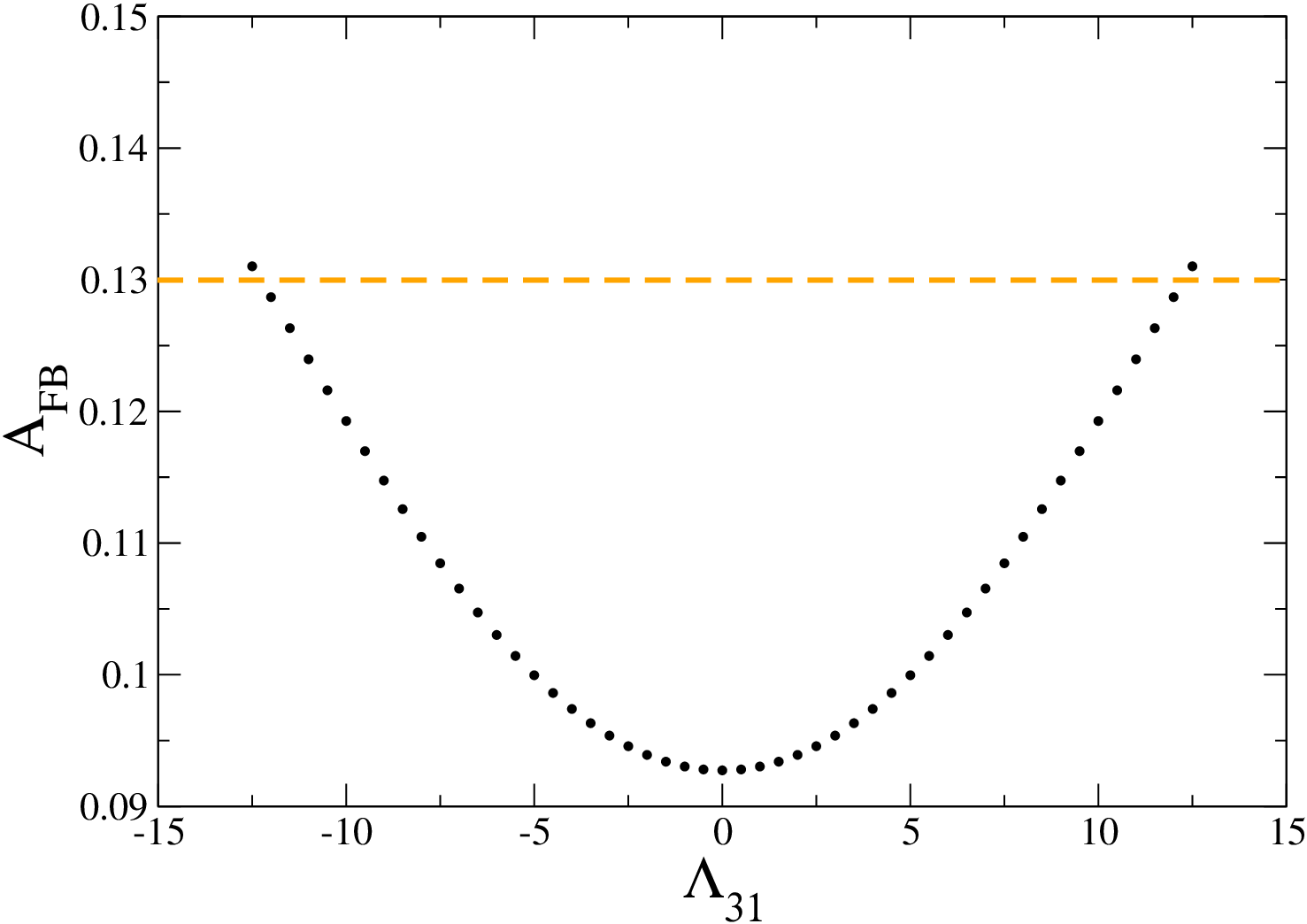}~~~
\includegraphics[width=2.8in]{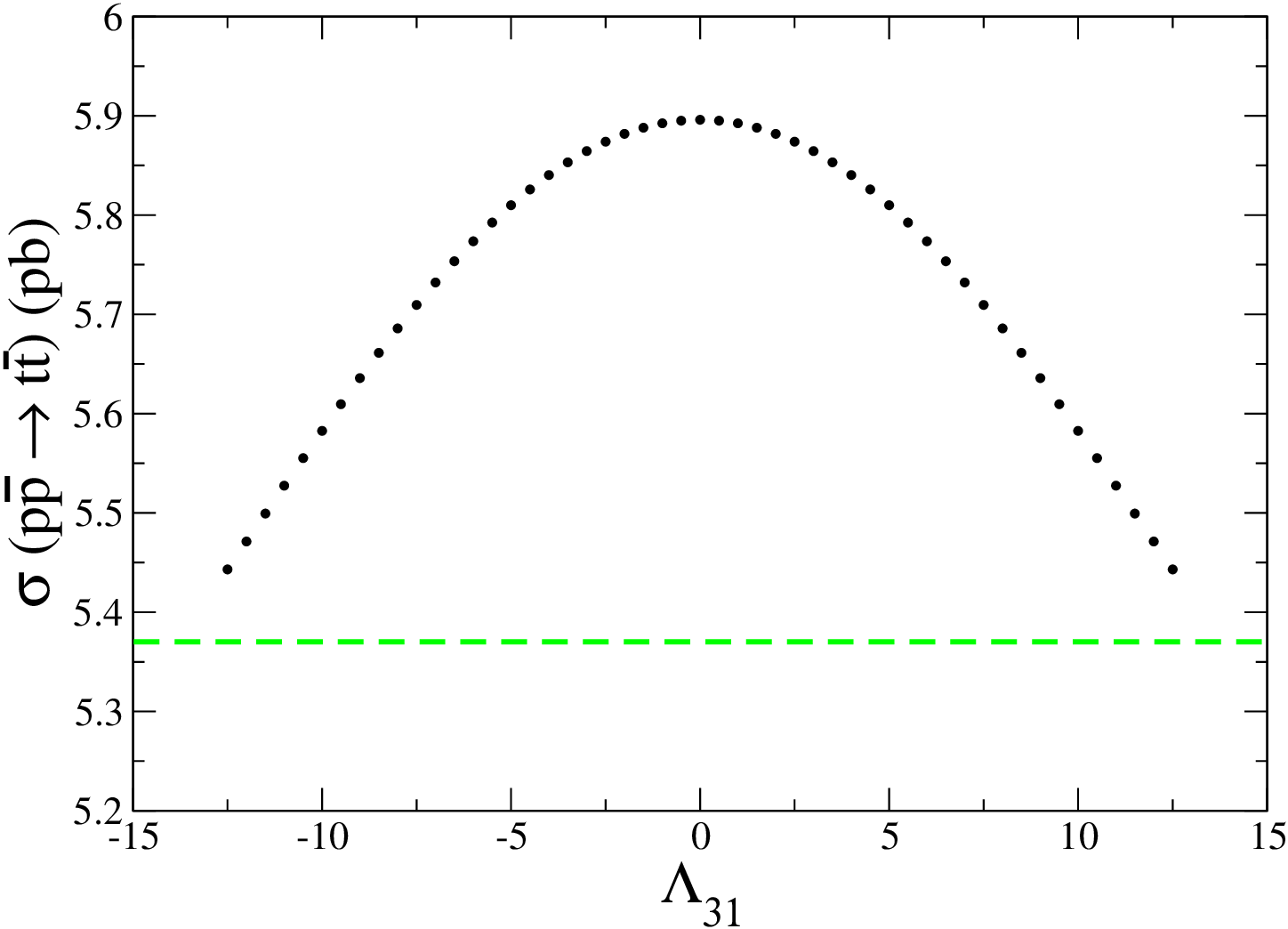}
\caption{\small  On the left plot the forward-backward top asymmetry at the parton level as a function of $\Lambda_{31}$ for the large $\mu_{s}$ scenario. The orange line indicates a one $\sigma$ deviation from a combination of the independent CDF and D$\O$ asymmetry measurements~\cite{Aaltonen:2011kc,Abazov:2011rq}. On the right, the $t\bar{t}$ production cross section as a function of $\Lambda_{31}$. The green line corresponds to three $\sigma$ deviations away from the experimental cross section~\cite{ttbarCSCDF}.
}
\label{fig:largemus1}
\end{figure}

In the large $\mu_{s}$ limit, the singlet decouples from the theory and in the Higgs decoupling limit the only light scalar is the SM-like Higgs. Furthermore, within this class of models $v_{s}\to0$ and the most dominant contribution to the cross section and asymmetry arises from the coupling of the SM-like Higgs to the up and top quarks which is proportional to
 \begin{equation}
 \frac{\left(\Lambda_{13,31}\right)v\sin\beta}{M}O^{H}_{1,S}.
 \end{equation}
  The value of $O^{H}_{1,S}$ is very small since the SM-like Higgs has a very little singlet component, hence the additional suppression. In the analysis, we fix the $\mu$ parameter to be consistent with searches of supersymmetric particles carried out by LEP~\cite{PDG}. Our main results are shown in Figure~\ref{fig:largemus1}. On the left we have plotted the total $t\bar{t}$ cross section and on the right the top forward-backward asymmetry as a function of the $\Lambda_{31}$ while fixing the value of $\Lambda_{13}=12.5$. For this figure we have chosen $\mu_{s}=1.5$ TeV, $\mu=500$ GeV, $A_{\lambda}=-1$ TeV and $B_{\mu}=(500~\text{GeV})^{2}$ which yield a value of $v_{s}=0.5$ GeV. We can see from the figure that even for rather large values of both $\Lambda_{13}$ and $\Lambda_{31}$, the interference contribution to the cross section always dominates. This is due to the additional suppression in the coupling of the SM-like Higgs to the up and top quarks, see Equation (5.2). Furthermore, an asymmetry above $13$$\%$, that is within one sigma of the experimental result, can only be obtained when maximizing both $\Lambda_{13}$ and $\Lambda_{31}$. However, the corresponding cross section is close to being outside the three sigma region. Models with large $\mu_{s}$ and with only a relatively light scalar with SM-like couplings present a large amount of tension in the sense that in order to minimize the negative interference contributions to the cross section, one must sacrifice obtaining a large asymmetry. 
%
%
%
%

%
%
\section{Conclusions}
We have incorporated dimension-five operators to the S-MSSM that couple first and third generation quarks to scalars. We have studied their contributions to the $t\bar{t}$ production cross section and the forward-backward asymmetry. We have studied the two limiting cases of the S-MSSM that provide a natural solution to the Little Hierarchy and analyzed the effects that the distinct spectra have on mediating the new contributions to the $t\bar{t}$ cross section and asymmetry. 

We found that in the small $\mu_{s}$ limit we are able to generate an inclusive asymmetry consistent with the combined CDF and D$\O$ result~\cite{Aaltonen:2011kc,Abazov:2011rq} for values of $\Lambda_{31}<4\pi$ while being consistent with the experimental $t\bar{t}$ production cross section~\cite{ttbarCSCDF}. The relevant couplings in the Lagrangian are $O\left(1\right)$, and there exists regions where our effective theory approach still holds. Of course a more careful analysis incorporating higher dimensional operators will be interesting and it is left for future work. In essence this limiting case of the S-MSSM is consistent with what was found in an extension to the SM using light weak doublet scalars~\cite{Blum:2011fa}. The main difference is that our model is supersymmetric with an spectrum fixed by the Higgs sector of the S-MSSM. In addition, flavor constraints mediated by charged scalars are less stringent given that charged Higgses are decoupled in this kind of models~\cite{CDOP1,CDP2}.

In the large $\mu_{s}$ limit we found that in order to minimize the interference contributions to the cross section, we had to sacrifice the production of a large asymmetry. The best case scenario was when both $\Lambda_{13}$ and $\Lambda_{31}$ were rather large. This region of our model generates a value of the cross section close to laying outside three sigma from the experimental value. Furthermore in this region of parameter space, our couplings are too large that one is now within a non-perturbative regime. Because of this and the tension that arises in generating a large enough asymmetry while being consistent with experimental $t\bar{t}$ production cross section, the case with large $\mu_{s}$ is not as promising as a new physics scenario.


To conclude, we have shown that the S-MSSM with additional dimension-five operators coupling first and third generation of quarks to scalars provides an explanation for the anomaly on the Tevatron inclusive forward-backward top asymmetry within a supersymmetric scenario. We found that the small $\mu_{s}$ limit of the S-MSSM appears to be the most promising.

%
%
\section*{Acknowledgements}
%
%
%
A very special thanks to J. de Blas for useful discussions. I will also like to thank W. Altmannshofer, A. Delgado, S. Gori and C. Kolda for providing with very essential and important feedback regarding all aspects of this work. This work was supported in part by the Fermilab Fellowship in Theoretical Physics. Fermilab is operated by Fermi Research Alliance, LLC, under Contract DE-AC02-07-CH11359 with the US Department of Energy.

%
%
%

\end{document}